\documentclass[aps,preprint,superscriptaddress,amssymb]{revtex4}
\usepackage{amsmath,amssymb,setspace,graphicx,subfigure,epsfig}

\usepackage{amsfonts}

\usepackage{latexsym}

\usepackage{epsfig}

\usepackage{color}
\usepackage{subfigure}

\newcommand\lsim{\mathrel{\rlap{\lower4pt\hbox{\hskip1pt$\sim$}}
    \raise1pt\hbox{$<$}}}
\newcommand\gsim{\mathrel{\rlap{\lower4pt\hbox{\hskip1pt$\sim$}}
    \raise1pt\hbox{$>$}}}
\def\bea{\begin{eqnarray}}
\def\eea{\end{eqnarray}}
\def\ba{\begin{array}}
\def\ea{\end{array}}
\def\bc{\begin{center}}
\def\ec{\end{center}}

\begin{document}

\title{Wrong-Higgs Interactions without Flavor Problems and Their Effects on Physical Observables}

\author{Kyu Jung Bae}
\affiliation{Department of Physics, Seoul National University, Seoul, 151-747, Korea}

\date{\today}

\begin{abstract}
We consider the wrong-Higgs interactions such as type-III two Higgs doublet models.
Generally, such interactions cause flavor problems.
However, if new Yukawa interactions have the same flavor structure as that of the
standard model(SM), we do not have any flavor problems.
In this work we propose a microscopic model for the wrong-Higgs interactions
aligned with SM ones in the context of supersymmetry(SUSY) and show their phenomenological implications.
Low energy contraints from muon $g-2$ and rare $B$ decays can be relieved
and it can be viable to have low mass superparticle spectra with light dark matter
which is preferred by recent experiments such as DAMA/LIBRA, CDMS-II and CoGeNT.
We also briefly discuss modification of Higgs decay in colliders.
\end{abstract}

\maketitle

\section{introduction}
New physics above the electroweak scale is strongly required to be responsible for
electroweak symmetry breaking.
If new physics includes some particles with mass above the electroweak scale, their flavor 
structure has great effects on low energy phenomena by quantum corrections.
On the opposite way, we get a sense of properties of new physics particles by
observing low energy phenomena such as rare $B$ decays.

Instead of considering new particles, we can think some new interactions on the standard
model(SM) Yukawa sector.
For example, in type-III two Higgs doublet model(2HDM-III) we have generalized Yukawa
interactions \cite{Atwood:1996vj}
\begin{equation}
\begin{split}
-{\cal L}_{\text{Y}}=&-h_uH_uQu^c+h_dH_dQd^c+h_eH_dLe^c\\
&+h_u^{\prime}H_d^{\dagger}Qu^c+h_d^{\prime}H_u^{\dagger}Qd^c+h_e^{\prime}H_u^{\dagger}Le^c+\text{h.c.}
\end{split}
\end{equation}
where $AB=\epsilon_{ij}A^iB^j$, $A^{\dagger}B=A^{i*}B^i$ for doublets $A$ and $B$.
Here we suppress flavor, color and weak indices.
The new wrong-Higgs couplings $h_u^{\prime}$, $h_d^{\prime}$ and $h_e^{\prime}$
generally have different flavor structure from those of ordinary couplings $h_u$, $h_d$ and $h_e$.
In that case wrong-Higgs interactions cause flavor changing neutral current(FCNC) problems
and large lepton flavor violating(LFV) decays at the tree-level \cite{Atwood:1996vj}.
However, if $h_u^{\prime}$, $h_d^{\prime}$ and $h_e^{\prime}$ have the same flavor structure
as $h_u$, $h_d$ and $h_e$, {\it i.e.} $h_u^{\prime}, h_d^{\prime}, h_e^{\prime}
\propto h_u,h_d,h_e$, they are simultaneously diagonalized with ordinary Yukawa couplings
and make no additional flavor problems.

Nevertheless, they affect enormously low energy phenomena since their $\tan\beta$-behaviors
become different from that of usual two Higgs doublet model type-I or type-II cases 
because of the wrong-Higgs interations. Here $\tan\beta=\langle H_u\rangle/\langle H_d\rangle$ 
is ratio of Higgs vacuum expectation values(vevs).
Low energy observables such as muon $g-2$ and rare $B$ decays have 
Yukawa coupling dependence on their new physics contributions and can be either
enhanced or suppressed depending on choice of the parameter space.
This has important implications for low energy supersymmetry models.
It is well-known that low energy supersymmetry is one of the most promising candidates of beyond SM because SUSY resolves
hierarchy problem in Higgs scalar potential and naturally explains the electroweak symmetry breaking
driven by radiative correction of top Yukawa sector.
However, low energy SUSY spectra are constrained by low energy observables.
The lighter SUSY particles are, the more natural theory we have but the severer low energy constraints we undergo.
This is tension between theoretical and experimental requirements in the model.
In order to relieve such tension, loop contributions on each observables from SUSY particles cancel out one another
so that very special parameter regions must be chosen.
Meanwhile, Models with squeezed SUSY spectra \cite{Dermisek:2006ey,Dermisek:2006qj,Bae:2007pa}
can achieve it more or less without special parameter choices.
If we have wrong-Higgs interactions, such tension can be relieved without special parameter choices 
because of the effects of wrong-Higgs interactions we mentioned above.
%$h-b-\bar{b}$ cooupling is also changed so that Higgs decay modes, Higgs discovery
%channel and mass bound from experiments are modified.

Such wrong-Higgs interactions are generated by SUSY one-loop corrections \cite{Haber:2007dj,Dobrescu:2010mk}
but there is no reason that loop-induced wrong-Higgs couplings are proportional to ordinary Yukawa couplings.
In this work, we will show a microscopic model for natural wrong-Higgs couplings 
aligned with ordinary Yukawa couplings in the context of beyond minimal supersymmetric
standard model(BMSSM).
Further, in our case, wrong-Higgs interactions are generated in tree-level by integrating out heavy states
and can be much larger than those from loop corrections.
Therefore, our wrong-Higgs interactions can relieve aforementioned tension more effectively than loop generated ones.
We will show the wrong-Higgs Yukawa effects on low energy observables and their
implications on low mass SUSY spectra with light dark matter
which is preferred by recent experiments such as DAMA/LIBRA \cite{Bernabei:2010mq},
CDMS-II \cite{Ahmed:2009zw} and CoGeNT \cite{Aalseth:2010vx}.
Higgs decay modification in colliders are also briefly discussed.

\section{A microscopic model}
We consider two pairs of Higgs doublets, one pair is heavy and the other is light.
By some discrete symmetry heavy pair couples to quarks and leptons but light pair does not.
At low energy scale much lighter than heavy Higgs doublet, discrete symmetry breaks
and mixings between heavy pair and light pair are generated.
Integrating out heavy pair we will obtain ordinary and wrong-Higgs 
interactions between light pair and quarks and leptons.

In SUSY context, we have two pairs of Higgs doublet superfields,
$H_1$, $H_2$, $H_d$ and $H_u$.
There is a charge assignment of both Higgs pairs and matter fields under 
discrete $Z_2\times Z_2^{\prime}$ symmetry in Table \ref{charges}.
\begin{table}[thb]
\begin{center}
\begin{tabular}{|c|c|c|c|c|c|c|c|c|c|}
\hline
& $H_1$ & $H_2$ & $H_d$ & $H_u$ & $Q$ & $u^c$ & $d^c$ & $L$ & $e^c$\\
\hline\hline
$Z_2$ & $+$ & $+$ & $-$ & $+$ & $+$ & $+$ & $+$ & $+$ & $+$\\
$Z_2^{\prime}$ & $+$ & $+$ & $+$ & $-$ & $+$ & $-$ & $+$ & $+$ & $+$\\
\hline 
\end{tabular}
\caption{Charges under $Z_2\times Z_2^{\prime}$.}\label{charges}
\end{center}
\end{table}
From the charge assignment, K\"ahler potential and superpotential are given by
\begin{eqnarray}
K&=&H_1^{\dagger}H_1+H_2^{\dagger}H_2+H_d^{\dagger}H_d+H_u^{\dagger}H_u
+\text{matter fields},\\
W&=&-MH_1H_2-h_uH_uQu^c+h_dH_1Qd^c+h_eH_1Le^c.
\end{eqnarray}
Here we count out gauge superfields in K\"ahler potential since 
they produce 4-leg or 5-leg operators of Higgses(Higgsinos), 
gauge bosons(gauginos) and fermions(sfermions) after integrating out heavy pair.
They are irrelevant for our discussions except Higgs-gaugino-fermion-sfermion interactions, 
which will be discussed later.
It is noteworthy that we assign charges to make heavy pair couple to down-type quarks and leptons
and light pair couple up-type quarks.
This is because Yukawa couplings of heavy pair are $1/M$-suppressed to be ordinary ones
of light pair as will be shown.
Top quark Yukawa coupling is already $\cal O$(1) so that if heavy pair couples to top quark,
its Yukawa coupling must be $\cal O$(10) or more and we cannot deal with this perturbatively.
Even though we have wrong-Higgs top Yukawa coupling, it is suppressed by $\cot\beta$ because of the smallness
of down-type Higgs vev and nothing interesting happens.
Therefore we consider the wrong-Higgs interactions in only down-type quark sector and lepton sector.

At the low energy scale, this $Z_2\times Z_2^{\prime}$ is supposed to be broken somehow.
Superpotential becomes
\begin{equation}
\begin{split}
W=&-MH_1H_2-\mu_{12} H_dH_u-\mu_1 H_1H_u-\mu_2 H_dH_2\\
&-h_uH_uQu^c+h_dH_1Qd^c+h_eH_1Le^c,\label{bmssm:W}
\end{split}
\end{equation}
where $M\sim \cal O$(1) TeV, $\mu_{12}, \mu_1, \mu_2 \sim \cal O$(100) GeV are
order parameters of discrete symmetry breaking such as quark masses in
chiral perturbation theory of strong interaction.
Integrating out heavy Higgs pair $H_1$ and $H_2$, K\"ahler potential and superpotential become
\begin{eqnarray}
K&=&\biggl(1+\biggl|\frac{\mu_2}{M}\biggr|^2\biggr)H_d^{\dagger}H_d
+\biggl(1+\biggl|\frac{\mu_1}{M}\biggr|^2\biggr)H_u^{\dagger}H_u+\text{four-Fermi interations}\nonumber\\
&&-\frac{\mu_1^*}{M^2}H_u^{\dagger}(h_dQd^c+h_eLe^c)+\text{h.c.},\\
W&=&-\biggl(\mu_{12}-\frac{\mu_1\mu_2}{M}\biggr)H_dH_u-h_uH_uQu^c
-\frac{\mu_2}{M}H_d(h_dQd^c+h_eLe^c)\nonumber\\
&&+\frac{1}{M}h_uh_eQu^cLe^c
\end{eqnarray}
Redefining parameters as
\begin{eqnarray}
&\mu_{12}-\frac{\mu_1\mu_2}{M}\equiv\mu, 
\qquad -\frac{\mu_2}{M}(h_d, h_e) \to (h_d, h_e),
\qquad \frac{1}{M}\big(\frac{\mu_1}{\mu_2}\big)\to\frac{1}{M}\label{para}
\end{eqnarray}
and neglecting five-dimensional matterfield interaction terms,
we obtain superpotential of light Higgs pair and matter fields, effectively.
Moreover, we have new wrong-Higgs interactions in K\"ahler potential 
such as Dine, Seiberg and Thomas \cite{Dine:2007xi}.

\section{Effective Potential}
\begin{figure}[thb]
\subfigure[Ordinary Yukawa interaction]{
\includegraphics[width=6cm]{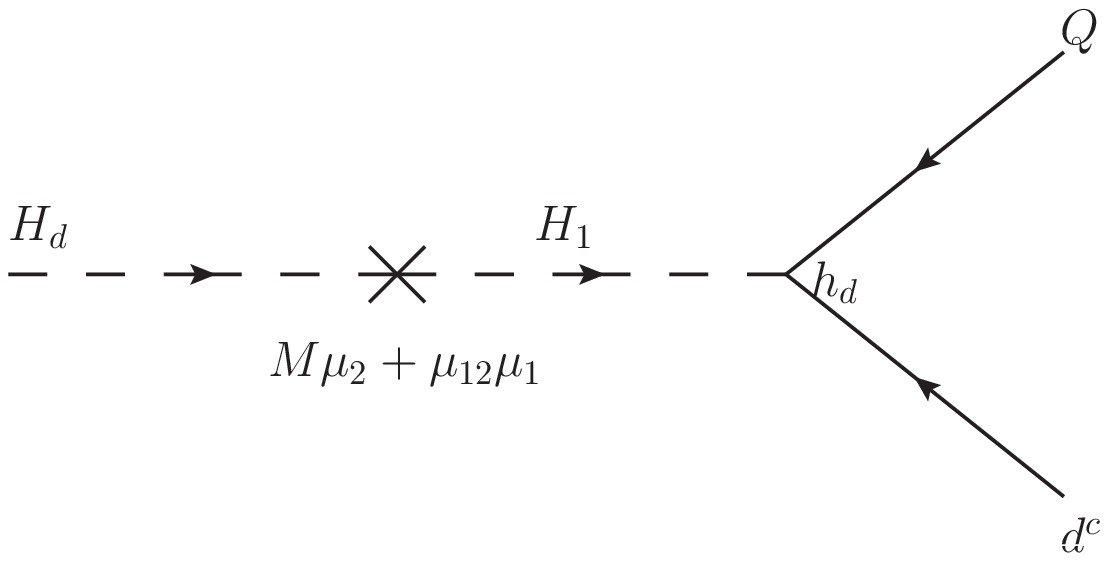}
\label{fig:feyn1}}
\qquad
\subfigure[Wrong-Higgs interaction]{
\includegraphics[width=6cm]{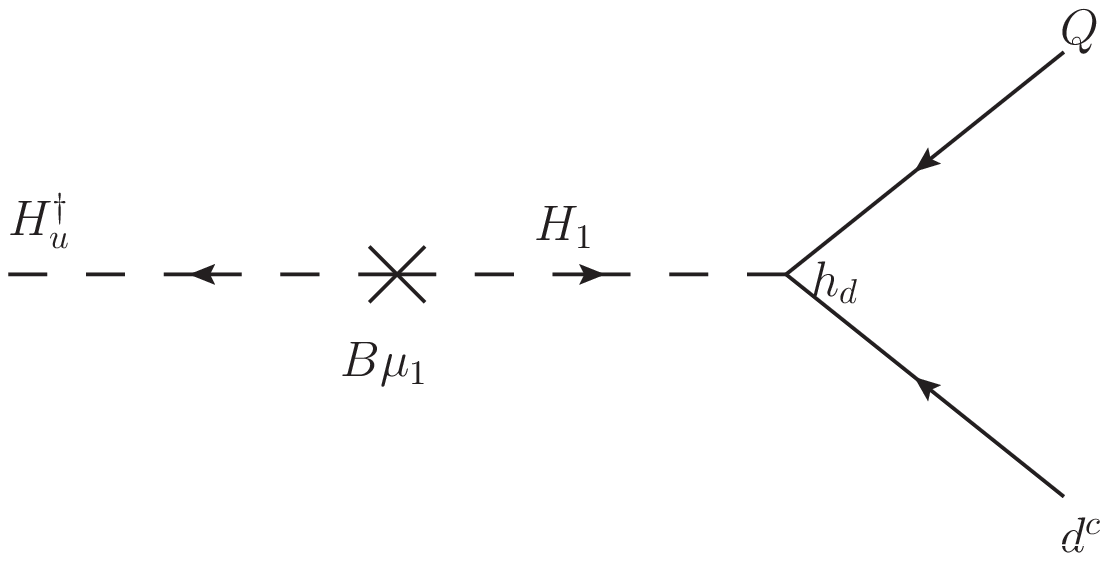}
\label{fig:feyn2}}
\caption{Feynman diagrams contributing ordinary and wrong-Higgs interactions}
\end{figure}

Including SUSY breaking spurion field ${\cal Z}=m_S\theta^2$, we obtain \cite{Dine:2007xi}
\begin{equation}
\int d^4 \theta \frac{1}{M}(1+{\cal Z+Z^{\dagger}+ZZ^{\dagger}})(h_dH_u^{\dagger}Qd^c+h_eH_u^{\dagger}Le^c).
\label{bmssm:K}
\end{equation}
From the above potential we have
\begin{eqnarray}
\delta V^{\text{K\"ahler}}_d&=&-\frac{h_d}{M}\big(h_u^2|\tilde{u}^c|^2\tilde{d}^cH_u^{\dagger}\tilde{Q}
-h_uh_d\tilde{u}^c|\tilde{d}^c|^2H_d^{\dagger}\tilde{Q}+h_u\tilde{u}^c\tilde{Q}Qd^c\nonumber\\
&&-h_u\mu\tilde{u}^{c*}\tilde{d}^cH_u^{\dagger}H_d+h_d\mu|\tilde{d}^c|^2H_d^{\dagger}H_d
-h_d\mu|H_d\tilde{Q}|^2-\mu H_dQd^c\big)+\text{h.c.}\nonumber\\
&&-\frac{h_dm_{{S}}}{M}\mu H_d\tilde{Q}\tilde{d}^c+\text{h.c.}\nonumber\\
&&-\frac{h_dm_{{S}}^*}{M}\big(h_d|\tilde{d}^c|^2H_u^{\dagger}H_d^{\dagger}
-h_dH_u^{\dagger}\tilde{Q}(H_d\tilde{Q})^*-H_u^{\dagger}Qd^c\big)+\text{h.c.}\nonumber\\
&&-\frac{h_d|m_{{S}}|^2}{M}H_u^{\dagger}\tilde{Q}\tilde{d}^c+\text{h.c.},\\
\delta V^{\text{K\"ahler}}_e&=&\frac{h_e}{M}\big(h_uh_e\tilde{u}^c|\tilde{e}^c|^2H_d^{\dagger}\tilde{Q}
-h_uh_e\tilde{u}^c\tilde{Q}\tilde{L}(H_d\tilde{L})^*-h_u\tilde{u}^c\tilde{Q}Le^c\nonumber\\
&&-h_e\mu|\tilde{e}^c|^2H_d^{\dagger}H_d+h_e\mu|H_d\tilde{L}|^2+\mu H_dLe^c\big)+\text{h.c.}\nonumber\\
&&+\frac{h_em_{{S}}}{M}\big(h_u\tilde{u}^c\tilde{e}^c\tilde{Q}\tilde{L}
-\mu H_d\tilde{L}\tilde{e}^c\big)+\text{h.c.}\nonumber\\
&&-\frac{h_em_{{S}}^*}{M}\big(h_e|\tilde{e}^c|^2H_u^{\dagger}H_d^{\dagger}
-h_eH_u^{\dagger}\tilde{L}(H_d\tilde{L})^*-H_u^{\dagger}Le^c\big)+\text{h.c.}\nonumber\\
&&-\frac{h_e|m_{{S}}|^2}{M}H_u^{\dagger}\tilde{L}\tilde{e}^c+\text{h.c.}.
\end{eqnarray}
Keeping only the leading contributions, we get
\begin{equation}
\begin{split}
\delta V^{\text{K\"ahler}}=
&\big(h_d\eta_{\mu}H_dQb^c-h_d\eta_SH_u^{\dagger}Qd^c+h_d\eta_S\mu H_d\tilde{Q}\tilde{d}^c
+h_d\eta_Sm_{{S}}H_u^{\dagger}\tilde{Q}\tilde{d}^c\big)+\text{h.c.}\\
&+\big(h_e\eta_{\mu}H_dLe^c-h_e\eta_SH_u^{\dagger}Le^c
+h_e\eta_S\mu H_d\tilde{L}\tilde{e}^c
+h_e\eta_Sm_SH_u^{\dagger}\tilde{L}\tilde{e}^c\big)+\text{h.c.}\\
&+\text{higher order couplings},
\end{split}\label{pot:kahler}
\end{equation}
where $\eta_{\mu}=\mu/M$ and $\eta_S=-m_S/M$.
As shown in eq. (\ref{pot:kahler}), wrong-Higgs interactions actually come from
SUSY breaking effect, which is proportional to $\eta_S$.
The reason is the following.
The first term of K\"ahler potential (\ref{bmssm:K}) comes from mixing term between
Heavy Higgs and light Higgs in superpotential (\ref{bmssm:W}).
Before integrating out we have the following $F$-term potential,
\begin{equation}
\begin{split}
V \supset& |F_{H_2}|^2+|F_{H_u}|^2\\
=&|MH_1+\mu_2H_d+\cdots|^2+|\mu_{12}H_d+\mu_1H_1+\cdots|^2\\
=&(M\mu_2+\mu_{12}\mu_1)H_1^{\dagger}H_d+\text{h.c.}+\cdots.
\end{split}
\end{equation}
From the first term in the third line, we have ordinary Yukawa interactions in superpotential 
after integrating out heavy pair and parameter redefinition (\ref{para})
and the second term becomes small correction ordinary Yukawa interactions 
proportional to $\eta_{\mu}$, see Fig. \ref{fig:feyn1}.
On the other hand, wrong-Higgs interactions come from SUSY breaking $B$ term of
$H_u$ and $H_1$, see Fig. \ref{fig:feyn2}.

Consequently, relations between Higgs vev and Yukawa couplings are changed as
\begin{eqnarray}
&h_d=\frac{m_d}{v_d(1+\xi_d\tan\beta)}, \qquad &\xi_d=-\eta_S+\eta_{\mu}\cot\beta,\label{yukawa:d}\\
&h_e=\frac{m_e}{v_d(1+\xi_e\tan\beta)}, \qquad &\xi_e=-\eta_S+\eta_{\mu}\cot\beta.\label{yukawa:e}
\end{eqnarray}
Down-type squark and slepton mass matrices are also changed as follows,
\begin{equation}
\begin{split}
m_{\tilde{d}}^2=&
\begin{pmatrix}
m_{\tilde{d}_L}^2 & h_dv_d(A_d^*-\mu\tan\beta) \\ h_dv_d(A_d-\mu^*\tan\beta) & m_{\tilde{d}_R}^2
\end{pmatrix}\\
&+
\begin{pmatrix}
0 & h_dv_d(\eta_S\mu^*+\eta_Sm_S\tan\beta)\\ h_dv_d(\eta_S\mu+\eta_Sm_S\tan\beta) & 0
\end{pmatrix}
\end{split}
\end{equation}
\begin{equation}
\begin{split}
m_{\tilde{e}}^2=&
\begin{pmatrix}
m_{\tilde{e}_L}^2 & h_{e}v_d(A_{e}^*-\mu\tan\beta) \\ h_{e}v_d(A_{e}-\mu^*\tan\beta) & m_{\tilde{e}_R}^2
\end{pmatrix}\\
&+
\begin{pmatrix}
0 & h_{e}v_d(\eta_S\mu^*+\eta_Sm_S\tan\beta)\\ h_{e}v_d(\eta_S\mu+\eta_Sm_S\tan\beta) & 0
\end{pmatrix}.
\end{split}
\end{equation}

\section{Muon $g-2$}

In the limit of large $\tan\beta$, SUSY contributions to muon $g-2$ consist of
four neutralino-smuon loops and one chargino-sneutrino loop \cite{Moroi:1995yh}.
For such case, each contribution is calculated by Moroi \cite{Moroi:1995yh}.
It is noticeable that each contribution proportional to $m_{\mu}^2$.
One power comes from equation of motion of free muon by the definition of $g-2$, and the other
comes from muon Yukawa coupling due to chirality flipping in the loop.
Hence muon $g-2$ is modified by Yukawa coupling corrections (\ref{yukawa:e}).
Including new Yukawa corrections, Moroi's formulae become
\begin{eqnarray}
\Delta a_{\mu}^{N1}&=&\frac{g_1^2m_{\mu}^2M_1(\mu-\eta_Sm_S)\tan\beta}{1+\xi_{\mu}\tan\beta}
\{J_5(M_1^2,M_1^2,m_{\tilde{\mu}L}^2,m_{\tilde{\mu}R}^2,m_{\tilde{\mu}R}^2)\nonumber\\
&&+J_5(M_1^2,M_1^2,m_{\tilde{\mu}R}^2,m_{\tilde{\mu}L}^2,m_{\tilde{\mu}L}^2)\},\label{mdm:N1}\\
\Delta a_{\mu}^{N2}&=&-\frac{g_1^2m_{\mu}^2M_1\mu\tan\beta}{1+\xi_{\mu}\tan\beta}
\{J_5(M_1^2,M_1^2,\mu^2,m_{\tilde{\mu}R}^2,m_{\tilde{\mu}R}^2)
+J_5(M_1^2,\mu^2,\mu^2,m_{\tilde{\mu}R}^2,m_{\tilde{\mu}R}^2)\},\label{mdm:N2}\\
\Delta a_{\mu}^{N3}&=&\frac12\frac{g_1^2m_{\mu}^2M_1\mu\tan\beta}{1+\xi_{\mu}\tan\beta}
\{J_5(M_1^2,M_1^2,\mu^2,m_{\tilde{\mu}L}^2,m_{\tilde{\mu}L}^2)
+J_5(M_1^2,\mu^2,\mu^2,m_{\tilde{\mu}L}^2,m_{\tilde{\mu}L}^2)\},\label{mdm:N3}\\
\Delta a_{\mu}^{N4}&=&-\frac12\frac{g_2^2m_{\mu}M_2\mu\tan\beta}{1+\xi_{\mu}\tan\beta}
\{J_5(M_2^2,M_2^2,\mu^2,m_{\tilde{\mu}L}^2,m_{\tilde{\mu}L}^2)
+J_5(M_2^2,\mu^2,\mu^2,m_{\tilde{\mu}L}^2,m_{\tilde{\mu}L}^2)\},\label{mdm:N4}\\
\Delta a_{\mu}^C&=&\frac{g_2^2 m_{\mu}^2 M_2 \mu \tan\beta}{1+\xi_{\mu}\tan\beta}
\{2I_4(M_2^2,M_2^2,\mu^2,m_{\tilde{\nu}}^2)-J_5(M_2^2,M_2^2,\mu^2,m_{\tilde{\nu}}^2,m_{\tilde{\nu}}^2)\nonumber\\\
&&+2I_4(M_2^2,\mu^2,\mu^2,m_{\tilde{\nu}}^2)-J_5(M_2^2,\mu^2,\mu^2,m_{\tilde{\nu}}^2,m_{\tilde{\nu}}^2)\}.\label{mdm:C}
\end{eqnarray}
where $\xi_{\mu}=\xi_e$ in this context, $N1$-$N4$ stand for the neutralino-smuon loops 
and $C$ stands for the chargino-sneutrino loop.
$I_N$ and $J_N$ are loop functions defined by \cite{Moroi:1995yh}
\begin{eqnarray}
I_N (m_1^2, \cdots, m_N^2) &=& \int \frac{d^4k}{(2\pi)^4 i}\frac{1}{(k^2-m_1^2) \cdots (k^2-m_N^2)},\\
J_N (m_1^2, \cdots, m_N^2) &=& \int \frac{d^4k}{(2\pi)^4 i}\frac{k^2}{(k^2-m_1^2) \cdots (k^2-m_N^2)}.
\end{eqnarray}

In the limit of large $\tan\beta$, $\xi_{\mu}\simeq -\eta_S$ and then all components of muon $g-2$, 
eqs. (\ref{mdm:N1})-(\ref{mdm:C}) are enhanced(suppressed) by $1/(1-\eta_S\tan\beta)$
for positive(negative) $\eta_S$.

\begin{figure}[thb]
\includegraphics[width=6cm]{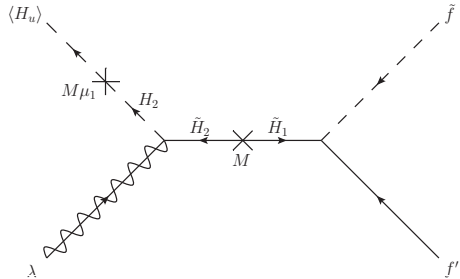}
\caption{Feynman diagram contributing gaugino-sfermion-fermion interactions}
\label{fig:feyn3}
\end{figure}

In addition to the leading contributions, there are corrections from gauge interactions.
In previous sections, we neglected gauge superfields in the K\"ahler potential.
Among them, gaugino-Higgs-sfermion-fermion interactions can contribute the muon $g-2$.
Their Feynman diagram is given in Fig. \ref{fig:feyn3}.
In usual cases, there are also gaugino-sfermion-fermion interactions, 
but these additional interactions make difference from those of usual cases.
These interactions flip chirality of fermions and sfermions
{\it i.e. $\lambda \tilde{q}_L q^{\prime}_R$},
so that can contribute magnetic moment interactions.
However these are suppressed by $v_u/M$ and have $\cal O$($\eta_{\mu}$) correction
for N2, N3, N4 and C diagrams of \cite{Moroi:1995yh}.
Since these do not have much effects on our calculations and conclusions of
large $\tan\beta$ behavior, we neglect them in our calculations.

\begin{figure}[thb]
\includegraphics[width=6cm]{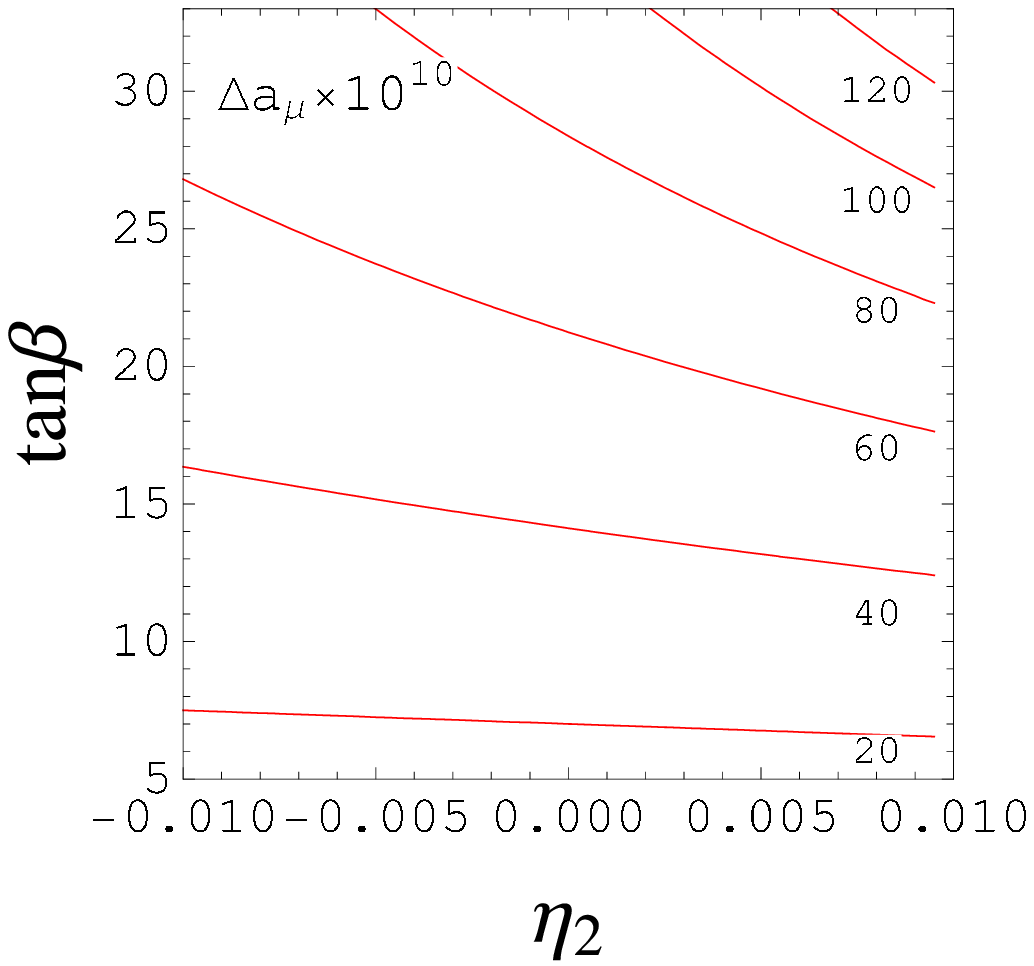}
\qquad
\includegraphics[width=6cm]{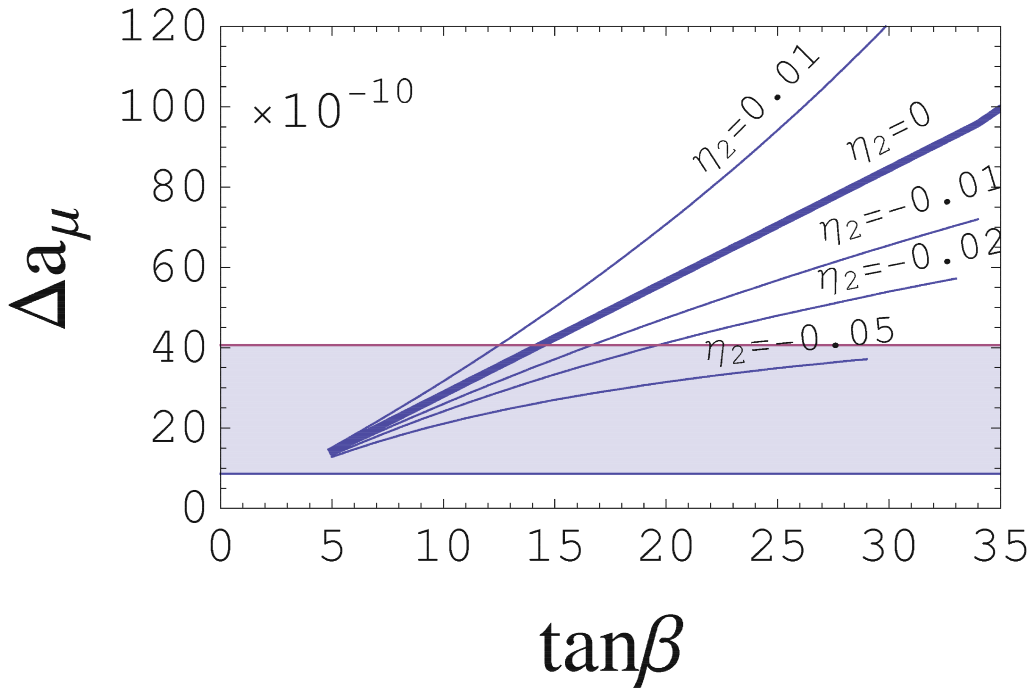}
\caption{$\Delta a_{\mu}$ for $m_0=100$ GeV, $M_{1/2}=250$ GeV, $A_0=-100$ GeV, $\mu>0$ and $\eta_{\mu}=0.01$
in constrained MSSM.
Left panel is contour plot of $\Delta a_{\mu}$ in $\eta_S-\tan\beta$ space and 
Right panel is $\Delta a_{\mu}$ for various $\eta_S$.
Shaded region in (b) is 2-$\sigma$ experimental bound \cite{Bennett:2006fi}. 
Numerical results are calculated by SuperIso v2.3 \cite{Mahmoudi:2008tp}.}
\label{fig:g-2}
\end{figure}

As seen in Fig. \ref{fig:g-2}, $\Delta a_{\mu}$ increases as $\tan\beta$ grows.
Futhermore, its growth become milder for negative $\eta_S$ and harder for positive $\eta_S$.
Strictly speaking, muon $g-2$ does not constrain models very much.
As seen in the right panel of Fig. \ref{fig:g-2}, models with low mass SUSY
spectra undergo muon $g-2$ tension in large $\tan\beta$ regime, though.
However, if $\eta_S=-0.05$, muon $g-2$ is still within 2-$\sigma$ experimental bound \cite{Bennett:2006fi}
for such large $\tan\beta\sim30$.
Muon $g-2$ tension is, therefore, relieved by wrong-Higgs correction.

\section{$B$ physics observables}
In this section we will show four representative $B$ observables 
such as ${\cal B}(B \to X_s \gamma)$, $\Delta M_{d,s}$, ${\cal B}(B_{d,s}\to\mu^+\mu^-)$ 
and ${\cal B}(B^{\pm}\to\tau^{\pm}\nu)$
which are affected by the wrong-Higgs interactions.
Basically, the wrong-Higgs interactions modify $\tan\beta$-behavior of down-type
quark Yukawa couplings.
Hence $B$ observables containing down-type quark Yukawa couplings are also modified in
the same direction as muon $g-2$ and $B$ physics constraints can be relieved.

We will show here the leading contributions for each observable.
The most dominant modifications come from relations between Yukawa couplings and Higgs vev (\ref{yukawa:d}).
In addition, new interaction $H_u^{\dagger}Qd^c$ can contribute Higgs mediated diagrams.
As will be discussed in next section, Higgs couplings to quarks are given by \cite{Djouadi:2005gj}
\begin{eqnarray}
g_{hdd}&=&-\frac{1}{\sqrt{2}}\frac{m_d}{v}\frac{\sin\alpha}{\cos\beta}
\biggl(\frac{1+\eta_{\mu}}{1+\eta_{\mu}-\eta_S\tan\beta}
+\frac{\eta_S\cot\alpha}{1+\eta_{\mu}-\eta_S\tan\beta}\biggr),\label{g_hdd}\\
g_{Hdd}&=&\frac{1}{\sqrt{2}}\frac{m_d}{v}\frac{\cos\alpha}{\cos\beta}
\biggl(\frac{1+\eta_{\mu}}{1+\eta_{\mu}-\eta_S\tan\beta}
-\frac{\eta_S\tan\alpha}{1+\eta_{\mu}-\eta_S\tan\beta}\biggr),\label{g_HHdd}\\
g_{Add}&=&-i\frac{1}{\sqrt{2}}\frac{m_d}{v}\tan\beta
\biggl(\frac{1+\eta_{\mu}}{1+\eta_{\mu}-\eta_S\tan\beta}
-\frac{\eta_S\cot\beta}{1+\eta_{\mu}-\eta_S\tan\beta}\biggr)\gamma_5,\label{g_Add}\\
g_{H^-u\bar{d}}&=&\frac{m_d}{v}V_{ud}\tan\beta
\biggl(\frac{1+\eta_{\mu}}{1+\eta_{\mu}-\eta_S\tan\beta}
-\frac{\eta_S\cot\beta}{1+\eta_{\mu}-\eta_S\tan\beta}\biggr)\frac12(1-\gamma_5)\nonumber\\
&&-\frac{m_u}{v}V_{ud}\cot\beta\frac12(1+\gamma_5),\label{g_H^-ud}\\
g_{H^+\bar{u}d}&=&\frac{m_d}{v}V_{ud}^*\tan\beta
\biggl(\frac{1+\eta_{\mu}}{1+\eta_{\mu}-\eta_S\tan\beta}
-\frac{\eta_S\cot\beta}{1+\eta_{\mu}-\eta_S\tan\beta}\biggr)\frac12(1+\gamma_5)\nonumber\\
&&-\frac{m_u}{v}V_{ud}^*\cot\beta\frac12(1-\gamma_5)\label{g_H^+ud},
\end{eqnarray}
where $\alpha$ is CP-even Higgs mixing angle, $V_{ud}$ is Cabibbo-Kobayashi-Maskawa(CKM) matrix
element, and $h$, $H$, $A$ and $H^{\pm}$ are two CP-even Higgses, CP-odd Higgs and charged Higgs.
In the regime of large $\tan\beta$, $|\tan\alpha|\simeq1/\tan\beta\ll1$ and second terms of 
eqs. (\ref{g_HHdd})-(\ref{g_H^+ud}) become negligible.
$H_u^{\dagger}Qd^c$ term affects our calculations only for light CP-even Higgs $h$.
Hence, $h$-mediated contribution is modified by new interactions as well as Higgs vev-Yukawa relation (\ref{yukawa:d}).
As we will see, however, $h$-mediated contributions are also negligible in our calculations. The reason is that down-type quarks are important for 
$B$ observables and down-type quarks mostly couple to heavy Higgs $H$ since heavy Higgs $H$ is mostly down-type for
large $\tan\beta$.
Detailed calculations and discussions are following.

\subsection{${\cal B}(B \to X_s \gamma)$}
SUSY contributions for ${\cal B}(B \to X_s \gamma)$ are composed of chargino-stop loop and charged Higgs-top loop.
In the large $\tan\beta$ regime, relevant contribution of chargino-stop loop
and charged Higgs-top loop to ${\cal B}(B\to X_s \gamma)$ are given by \cite{Carena:2000uj}
\begin{eqnarray}
{\cal B}(B\to X_s \gamma)|_{\tilde{\chi}^{\pm}}&\propto&\mu A_t\tan\beta \frac{m_b}{v(1+\Delta m_b)}
f(m_{\tilde{t}_1},m_{\tilde{t}_2},m_{\tilde{\chi}^{\pm}}),\\
{\cal B}(B\to X_s \gamma)|_{H^{\pm}}&\propto&
\frac{m_b(h_t\cos\beta-\delta h_t\sin\beta)}{v\cos\beta(1+\Delta m_b)}g(m_{H^{\pm}},m_t),
\end{eqnarray}
where $\delta h_t$ is supersymmetric QCD(SQCD) correction and
$f$ and $g$ are loop integrals.
$\Delta m_b$ is SQCD correction in ordinary MSSM but is the sum of SQCD correction and
wrong-Higgs correction (\ref{yukawa:d}) in our case, which is given by
\begin{equation}
\Delta m_b = (\xi_d+\epsilon_{bb})\tan\beta,
\end{equation}
where $\epsilon_{bb}$ is SQCD correction which is given in \cite{Gomez:2006uv}.
For large $\tan\beta$, chargino-stop contribution become dominant because it is 
proportional to $\tan\beta$ but charged Higgs-top contribution is not.
Furthermore, since the absolute value of $\Delta m_b$ increases as $\tan\beta$ grows,
$\tan\beta$-enhancement of ${\cal B}(B\to X_s \gamma)$ becomes milder(harder)
for negative(positive) $\Delta m_b$ as muon $g-2$.
In ordinary MSSM case, $\Delta m_b$ effects on ${\cal B}(B\to X_s \gamma)$
are studied in \cite{Carena:2000uj, Gomez:2006uv}.
Since interested parameter space in our case is for $0.01\lsim |\eta_S|\lsim0.1$,
wrong-Higgs correction is much larger than SQCD correction $\epsilon_{bb}\sim0.004$
and dominates SQCD correction.
Therefore, ${\cal B}(B\to X_s \gamma)$ constraint can be relieved in presence of 
wrong-Higgs correction($\eta_S<0$)
for low energy SUSY spectra such as muon $g-2$ constraint.

\subsection{$\Delta M_{d,s}$}
In supersymmetric models, $\Delta M_s$ is given by \cite{Buras:2002wq}
\begin{equation}
\Delta M_s=|(\Delta M_s)^{\text{SM}}+(\Delta M_s)^{H^{\pm}}+(\Delta M_s)^{\tilde{\chi}^{\pm}}
+(\Delta M_s)^{\text{DP}}|,
\end{equation}
where $(\Delta M_s)^{\text{SM}}$ is SM contribution, $(\Delta M_s)^{H^{\pm}}$ is charged Higgs
box-diagram contribution, $(\Delta M_S)^{\tilde{\chi}^{\pm}}$ is chargino box-diagram contribution,
and $(\Delta M_s)^{\text{DP}}$ is double penguin diagram contribution described in \cite{Buras:2002wq}.
For large $\tan\beta\gsim50$, $(\Delta M_s)^{\text{DP}}$ is a dominant contribution of ${\cal O}(\tan^4\beta)$,
which is given by \cite{Buras:2002wq}
\begin{equation}
\begin{split}
(\Delta M_s)^{\text{DP}}=&-12.0/\text{ps}\times\biggl[\frac{\tan\beta}{50}\biggr]^4
\biggl[\frac{P_2^{\text{LR}}}{2.5}\biggr]\biggl[\frac{F_{B_S}}{230\text{ MeV}}\biggr]^2
\biggl[\frac{|V_{ts}|}{0.040}\biggr]^2\\
&\times\biggl[\frac{\bar{m}_b(\mu_t)}{3.0 \text{ GeV}}\biggr]
\biggl[\frac{\bar{m}_s(\mu_t)}{0.06 \text{ GeV}}\biggr]
\biggl[\frac{\bar{m}_t^4(\mu_t)}{M_W^2M_A^2}\biggr]
\frac{\epsilon_Y^2(16\pi^2)^2}{(1+\tilde{\epsilon}_3\tan\beta)^2(1+\epsilon_0\tan\beta)^2},
\end{split}\label{dM_b}
\end{equation}
where $P_2^{\text{LR}}\approx2.5$ includes NLO QCD corrections and relavant hadronic matrix elements,
$F_{B_s}$ is $B_s$ decay constant, $M_A$ is CP-odd Higgs boson mass,
$\bar{m}_{b,s,t}(\mu_t)$ are effective running mass parameter of $b,s,t$-quark,
and $\epsilon_0$, $\epsilon_Y$ and $\tilde{\epsilon}_3$ 
come from one-loop SQCD correction defined in \cite{Buras:2002wq}.
It is noteworthy that these double penguin contributions are mediated by neutral Higgs bosons $h$, $H$ and $A$.
As discussed at the beginning of this section, wrong-Higgs interaction $H_u^{\dagger}Qd^c$ modify
light CP-even Higgs $h$ coupling to down-type quarks.
Although, CP-even Higgs constribution is proportional to $\sin^2(\alpha-\beta)\ll1$ so that is negligible
in decoupling limit with $M_A\gg M_Z$ and $\tan\beta\gg1$ where $M_Z$ is $Z$ boson mass.
On the other hand, heavy CP-even Higgs contribution is proportional to $\cos^2(\alpha-\beta)$ and
$M_H\approx M_A$ in the same limit so that contributions from $H$ and $A$ are almost the same \cite{Buras:2002wq}.
Hence we obtain the above result (\ref{dM_b}).
$\Delta M_d$ is given by the same formula with $\bar{m}_s$, $F_{B_s}$ and $V_{ts}$
replaced by $\bar{m}_d$, $F_{B_d}$ and $V_{td}$.

As ${\cal B}(B \to X_s \gamma)$ case, SQCD one-loop corrections are ${\cal O}(10^{-3})$
but wrong-Higgs correction $\eta_S$ is ${\cal O}(10^{-2}-10^{-1})$ in our interest.
Therefore, wrong-Higgs correction dominates SQCD corrections in ordinary MSSM and we can replace parameters as
\begin{equation}
\tilde{\epsilon}_3\rightarrow\tilde{\epsilon}_3+\xi_d\approx -\eta_S
\qquad \epsilon_0\rightarrow\epsilon_0+\xi_d\approx -\eta_S
\end{equation}
and
\begin{equation}
(\Delta M_{d,s})^{\text{DP}} \propto \frac{\tan^4\beta}{(1-\eta_S\tan\beta)^4}.
\end{equation}
In this case, we can make $\tan\beta$-dependence milder(harder) for
negetive(positive) $\eta_S$.
For $\tan\beta\lsim50$, charged Higgs box-diagram contribution is also substantial.
This diagram also have two down-type quark Yukawa couplings
and we have $1/(1-\eta_S\tan\beta)^2$ as well.

Since wrong-Higgs correction is universal for down-type quarks and leptons,
we have $1/(1-\eta_S\tan\beta)$ correction factor for each down type Yukawa coupling
in the whole interaction vertices or loop diagrams such as the aforementioned observables.

\subsection{${\cal B}(B_s\to\mu^+\mu^-)$}
In large $\tan\beta$ regime, ${\cal B}(B_s\to\mu^+\mu^-)$ can constrain our models.
For $\tan\beta\sim50$, ${\cal B}(B_s\to\mu^+\mu^-)$ is given by \cite{Buras:2002wq}
\begin{equation}
\begin{split}
{\cal B}(B_s\to\mu^+\mu^-)=&3.5\times10^{-5}\biggl[\frac{\tan\beta}{50}\biggr]^6
\biggl[\frac{\tau_{B_s}}{1.5\text{ ps}}\biggr]\biggl[\frac{F_{B_S}}{230 \text{ MeV}}\biggr]^2
\biggl[\frac{|V_{ts}^{\text{eff}}|}{0.040}\biggr]^2\\
&\times\frac{\bar{m}_t^4}{M_A^4}
\frac{\epsilon_Y^2(16\pi^2)^2}{(1+\tilde{\epsilon}_3\tan\beta)^2(1+\epsilon_0\tan\beta)^2},
\label{bsmumu}
\end{split}
\end{equation}
where $\tau_{B_s}$ is lifetime of $B_s$ meson and
$V_{ts}^{\text{eff}}$ is one-loop effective CKM matrix element.
This is also mediated by neutral Higgs bosons $h$, $H$ and $A$ such as $\Delta M_{d,s}$.
By the same reason discussed in the above subsection and the fact that
Higgs boson couplings to leptons have the same form of those of down-type quarks
except for their Yukawa coupling constants({\it i.e.} $H_u^{\dagger}Le^c$ can affect only $h$ couplings to leptons),
$h$-mediated contribution is negligible
and $H$- and $A$-mediated contribution dominate so we obtain (\ref{bsmumu}).
${\cal B}(B_d\to\mu^+\mu^-)$ is given by
\begin{equation}
\frac{{\cal B}(B_d\to\mu^+\mu^-)}{{\cal B}(B_s\to\mu^+\mu^-)}
=\biggl[\frac{\tau_{B_d}}{\tau_{B_s}}\biggr]\biggl[\frac{F_{B_d}}{F_{B_s}}\biggr]^2
\biggl[\frac{|V_{td}^{\text{eff}}|}{|V_{ts}^{\text{eff}}|}\biggr]^2
\biggl[\frac{M_{B_d}}{M_{B_s}}\biggr]^5.
\end{equation}
In this case, one-loop parameters $\epsilon_0$ and $\tilde{\epsilon}_3$ are replaced by
$-\eta_S$ as before.
We have lepton Yukawa coupling in (\ref{bsmumu}) so we have two more
factors of $1/(1-\eta_S\tan\beta)$ from lepton Yukawa coupling.
Hence we obtain
\begin{equation}
{\cal B}(B_{d,s}\to\mu^+\mu^-)\propto\frac{\tan^6\beta}{(1-\eta_S\tan\beta)^6}.
\end{equation}
If $\eta_S=-0.05$ and $\tan\beta=50$, we have $1/(3.5)^6$ suppression factor but $1/(1.25)^4$
in ordinary MSSM with $\epsilon_0\sim\tilde{\epsilon}_3\sim0.005$.
We have $\cal O$$(10^{-3})$ more suppression factor than in usual cases.

\subsection{${\cal B}(B^{\pm}\to\tau^{\pm}\nu)$}
In $B^{\pm}\to\tau^{\pm}\nu$ decay, SM contribution is from $W$ boson mediated diagram.
In MSSM cases, charged Higgs mediated diagram also has sizeable and destructive contribution.
For large $\tan\beta$ regime, MSSM contribution is given by \cite{Isidori:2006pk}
\begin{equation}
R_{B\tau\nu}=\frac{{\cal B}(B^{\pm}\to\tau^{\pm}\nu)_{\text{MSSM}}}
{{\cal B}(B^{\pm}\to\tau^{\pm}\nu)_{\text{SM}}}
=\biggl[1-\biggl(\frac{m_B^2}{m_{H^{\pm}}^2}\biggr)
\frac{\tan^2\beta}{1+\epsilon_0\tan\beta}\biggr]^2,
\end{equation}
where $m_B$ is $B$ meson mass and $m_{H^{\pm}}$ is charged Higgs boson mass.
For the same reason as before, $\epsilon_0$ is replaced by $-\eta_S$ and we have
one more $1/(1-\eta_S\tan\beta)$ factor from lepton Yukawa coupling.
Hence $R_{B\tau\nu}$ becomes
\begin{equation}
R_{B\tau\nu}
=\biggl[1-\biggl(\frac{m_B^2}{m_{H^{\pm}}^2}\biggr)
\frac{\tan^2\beta}{(1-\eta_S\tan\beta)^2}\biggr]^2,
\end{equation}
As the other low energy observables, $B^{\pm}\to\tau^{\pm}\nu$ decay also
have milder(harder) $\tan\beta$ dependence for negative(positive) $\eta_S$.

\section{Higgs decay}
In ordinary MSSM or SM, dominant decay mode of light Higgs is $b$-quark pair
unless Higgs mass is larger than twice the mass of $W$ boson \cite{Djouadi:2005gj, Djouadi:2005gi}.
In our case, interaction between Higss and $b$-quark is given by 
\begin{equation}
h_b(1+\eta_{\mu})H_d^0bb^c-h_b\eta_SH_u^{0*}bb^c.
\end{equation}
The physical CP-even Higgs bosons are obtained by 
\begin{equation}
\begin{pmatrix}
H\\h
\end{pmatrix}
=
\begin{pmatrix}
\cos\alpha & \sin\alpha \\ -\sin\alpha & \cos\alpha
\end{pmatrix}
\begin{pmatrix}
H_d^0\\H_u^0
\end{pmatrix},
\end{equation}
where $H$ and $h$ are eigenstates of CP-even Higgs mass matrix
\begin{equation}
\begin{pmatrix}
M_A^2\sin^2\beta+M_Z^2\cos^2\beta&-\frac12(M_A^2+M_Z^2)\sin2\beta\\
-\frac12(M_A^2+M_Z^2)\sin2\beta&M_A^2\cos^2\beta+M_Z^2\sin^2\beta
\end{pmatrix}.\label{mass:even}
\end{equation}
Using this relation and Yukawa relation (\ref{yukawa:d}), 
the lightest CP-even Higgs boson coupling to $b$-quark is given by
\begin{equation}
g_{hbb}=-\frac{1}{\sqrt{2}}\biggl(\frac{m_b}{v}\frac{\sin\alpha}{\cos\beta}\biggr)
\biggl(\frac{1+\eta_{\mu}+\eta_S\cot\alpha}{1+\eta_{\mu}-\eta_S\tan\beta}\biggr).
\label{vertex:hbb}
\end{equation}
In usual cases, Higgs mixing angle $\alpha$ is given by
\begin{equation}
\alpha=\frac12\biggl(\tan2\beta\frac{M_A^2+M_Z^2}{M_A^2-M_Z^2}\biggr),
\qquad -\frac{\pi}{2}\leq\alpha\leq0.
\end{equation}
In the limit of large $\tan\beta$, $\sin\alpha\simeq-\cos\beta$ and the term in the second
parenthesis of (\ref{vertex:hbb}) becomes unity so that $g_{hbb}$ is the same as that of 
ordinary MSSM case.
However, if there are some higher dimensional operators \cite{Dine:2007xi} 
or finite threshold corrections \cite{Kim:2009sy} in Higgs scalar potential, 
Higgs mixing angle $\alpha$ can be positive and Higgs collider phenomenology
can be modified \cite{Kim:2009sy, Bae:2010cd}.
According to \cite{Kim:2009sy}, Higgs mixing angle can be nearly zero 
in some special parameter region
in which a correction cancels a tree-level off-diagonal mass term of CP-even Higgs mass matrix (\ref{mass:even})
and Higgs mixing vanishes.
In this case Higgs decay to $b$-quark is highly suppressed 
and Higgs to $W$ boson decay channel dominates the others.
In our case, there is an additional modification factor in (\ref{vertex:hbb}).
For large $\tan\beta$ and large $M_A$ case, we can have two-fold suppression factor in (\ref{vertex:hbb}).
One is from $1/(1-\eta_S\tan\beta)$ factor for negative $\eta_S$ as before
and the other is from $(1+\eta_S\cot\alpha)$ factor.
If off-diagonal correction is of the same order of tree-level part and just bigger than that,
we have positive but small $\alpha$ because $M_A$ is much larger than $M_Z$ and $\tan\beta$ is large.
For example, if $\tan\beta\sim\cot\alpha\sim10$ and $\eta_S=0.05$,
we have 1/3 suppression of $g_{hbb}$ coupling from the relation (\ref{vertex:hbb})
and Higgs decay to $b$-quark cross section become 1/10 of that of usual cases.
Branching ratio of Higgs decay to $W$ bosons become larger and dominant.
Hence we have similar phenomenology to \cite{Kim:2009sy} without special parameter choices.

\section{Conclusions}
In this paper, we considered the phenomenological implication of
the wrong-Higgs interactions.
We proposed a microscopic model which give wrong-Higgs couplings
aligned with ordinary Yukawa couplings.

We showed that the wrong-Higgs interactions in down-type quark sector and
lepton sector generate different $\tan\beta$-dependence of relation 
between Higgs vev and Yukawa couplings.
Consequently, low energy observables involving down-type Yukawa couplings 
can be modified and their constraint on low energy SUSY spectra can be
relieved by negative $\eta_S$.
Futhermore, such relaxation can make it viable to explain recent dark matter 
experiments \cite{Bernabei:2010mq, Ahmed:2009zw, Aalseth:2010vx} by recently proposed dark matter
scenario by Kuflik {\it et al.} \cite{Kuflik:2010ah}.
In such scenario, we need very large $\tan\beta$ but this is strongly constrained
by $B_s\to \mu^+\mu^-$ decay.
In our situation, ${\cal B}(B_s\to \mu^+\mu^-)$ is suppressed by factor $\cal O$$(10^{-3})$
compared to the ordinary MSSM.
Another constraints are also relieved as well.

Finally, we showed the modification of Higgs decays in colliders.
In usual cases, Higgs decays predominantly to $b$-quark pair.
Under our circumstance, however, Higgs decay to $b$-quark pair can be suppressed
and $W$ boson channel become important for Higgs discovery.

\section*{Acknowledgement}
We thank Hyung Do Kim for useful discussion and comments.
This work is supported by KRF-2008-313-C00162 and CQUeST of Sogang University grant 2005-0049049.

\end{document}